\documentclass[pra,twocolumn,showpacs,superscriptaddress]{revtex4-1}
\usepackage[utf8]{inputenc}
\usepackage{graphicx}
\usepackage{amsmath}
\usepackage{esint}
\usepackage[dvipdfm, colorlinks=true, citecolor=blue, urlcolor=blue]{hyperref}

\begin{document}
\title{Exact decoherence-free state of two distant quantum systems in a non-Markovian environment}
\author{Chong Chen}
\affiliation{Center for Interdisciplinary Studies $\&$ Key Laboratory for Magnetism and Magnetic Materials of the MoE, Lanzhou University, Lanzhou 730000, China}
\author{Chun-Jie Yang}
\affiliation{Center for Interdisciplinary Studies $\&$ Key Laboratory for Magnetism and Magnetic Materials of the MoE, Lanzhou University, Lanzhou 730000, China}
\author{Jun-Hong An}\email{anjhong@lzu.edu.cn}
\affiliation{Center for Interdisciplinary Studies $\&$ Key Laboratory for Magnetism and Magnetic Materials of the MoE, Lanzhou University, Lanzhou 730000, China}

\begin{abstract}
Decoherence-free state (DFS) encoding supplies a useful way to avoid the detrimental influence of the environment on quantum information processing. The DFS was previously well established in either the two subsystems locating at the same spatial position or the dynamics under the Born--Markovian approximation. Here, we investigate the exact DFS of two spatially separated quantum systems consisting of two-level systems or harmonic oscillators coupled to a common non-Markovian zero-temperature bosonic environment. The exact distance-dependent DFS and the explicit criterion for forming the DFS are obtained analytically, which reveals that the DFS can arise only in one-dimensional environment. It is remarkable to further find that the DFS is just the system-reduced state of the famous bound state in the continuum (BIC) of the total system predicted by Wigner and von Neumann. On the one hand our result gives insight into the physical nature of the DFS, and on the other hand it supplies an experimentally accessible scheme to realize the mathematically curious BIC in the standard quantum optical systems.
\end{abstract}
\maketitle

\section{Introduction}
As a ubiquitous phenomenon in microscopic world, decoherence describes an inevitable loss of quantum coherence due to the interactions between quantum system and its environment. It is seen as a main obstacle to the realization of any applications utilizing quantum coherence, e.g., quantum computation \cite{Ladd2010}, quantum teleportation \cite{Bouwmeester1997}, and quantum metrology \cite{PhysRevLett.109.233601}. Therefore, how to control decoherence is a crucial issue in quantum engineering. Many active schemes, such as feedback control \cite{Wiseman1993} and dynamical decoupling \cite{Viola1999}, have been proposed to beat this unwanted effect. On the other hand, people found that decoherence can also be used for good purpose \cite{Diehl2008, Hansom2014, Kienzler2015}. It was found that the decoherence caused by a common environment can play a constructive role in generating stable entanglement between two quantum systems \cite{Plenio2002, Braun2002, Benatti2003, Lin2013}. The intrinsic physics is the existence of the decoherence-free state (DFS) \cite{Zanardi1997, Duan1997, Lidar1998}, which triggers the enthusiasm of relearning the role of decoherence of composite system caused by a common environment from different systems such as harmonic oscillators \cite{PhysRevA.76.042127, Horhammer2008, Zell2009, kajari2012, Voje2015} and spins \cite{CamposVenuti2006, Contreras-Pulido2008, Gonzalez-Tudela2011, Zheng2013}, and different environments such as crystal chains \cite{Wolf2011, Fogarty2013, Taketani2014} and waveguides \cite{Gonzalez-Tudela2011, Zheng2013,Gonzalez-Ballestero2013, Gonzalez-Ballestero2015, Zheng2013, Loo2013, Ramos2014}.

It is clear in principle that the DFS is present when the two quantum systems are at same spatial position \cite{Zanardi1997, Duan1997, Lidar1998}. While when they are spatially separated, there are still controversies on whether the DFS exists or not or, equivalently, whether the common environment can create stable entanglement distribution. Some works pointed out that the entanglement would disappear when the spatial distances are larger than the wavelength associated with the environmental cutoff frequency \cite{Audenaert2002, Solenov2006, Anders2008, Zell2009, Stauber2009}, while some other works claimed that the DFS for distant quantum systems is still possible to be formed \cite{Wolf2011, Zheng2013, Gonzalez-Ballestero2013, Fogarty2013, Taketani2014}. However, the physical nature of the DFS, especially its explicit form and its dependence on the spatial distance, and when it is formed have seldom been touched in these works. The realistic significance of answering these questions is that it could supply meaningful message for designing practical devices to distribute long-distance entanglement.

Another inspiration of our study is the famous bound state in the continuum (BIC), which was proposed soon after the birth of quantum mechanics \cite{Neumann}. Being stable in space but with its energy lying in the continuous energy band, such counterintuitive eigenstate of the quantum system was regarded as a mathematical curiosity due to the inaccessible potentials for a long time \cite{PhysRevA.11.446}. That situation changed when it was proposed that the BIC can arise naturally by virtue of the destructive interference between two resonance states in molecule system \cite{PhysRevA.32.3231}. Although the BIC has been extensively studied in the classical optical systems \cite{PhysRevLett.100.183902, PhysRevLett.107.183901, PhysRevLett.111.240403, Hsu.499.188, PhysRevLett.113.037401, PhysRevLett.113.257401, PhysRevLett.114.245503}, the BIC was rarely demonstrated in quantum systems.

In this work, we reveal that these two seemingly unrelated concepts merge together in open quantum systems. By studying the decoherence of two distant quantum systems consisting of either two-level systems (TLSs) or harmonic oscillators embedded in a common bosonic environment, we derive analytically the exact DFS and the physical criterion for forming the DFS without resorting to the Born--Markovian approximation (BMA). It is found that the distance-dependent DFS can only exist in a one-dimensional environment. This is  in sharp contrast with the case in which the two subsystems are located in the same spatial position, where the DFS is present irrespective of the environmental dimension. Further study reveals that the DFS, which scales as $1/R$ with increasing system distance $R$, corresponds exactly to the system reduced state of the BIC of the total system. The emergence of such a BIC can be physically attributed to the destructive interference between the two independent interaction channels of the two subsystems with the common environment, which can be seen as a direct realization of Friedrich and Wintgen's idea on the BIC \cite{PhysRevA.32.3231} in quantum optical system. Our conclusions are verified in the models of two TLSs interacting with a coupled cavity chain acting as an environment. Our study gives a realizable scheme to detect the BIC by observing the decoherence dynamics of open quantum systems.

The paper is organized as follows: In Sec. \ref{model}, we present our model. The DFS under and beyond the BMA is derived in Sec. \ref{DFSBMNBM}. The correspondence between the DFS and the BIC is also established here. By two examples of two TLSs interacting with the nearest-neighbor and the next-nearest-neighbor coupled-cavity arrays acting as environments, our conclusions are verified in Sec. \ref{illsmodel}. In Sec. \ref{con}, a summary is given.

\section{The Model}\label{model}
Consider two spatially separated quantum systems coupled to a common dissipative bosonic environment. The Hamiltonian reads $\hat{H}=\hat{H}_\text{S}+\hat{H}_\text{E}+\hat{H}_\text{I}$ with
\begin{align}\label{Hamiltonian}
\begin{split}
 \hat{H}_\text{S}&=\sum_{j=1,2}  \omega_0 \hat{O}_ {j} ^{\dag}\hat{O}_{j}, ~~\hat{H}_\text{E}= \sum_\mathbf{k} \omega_k \hat{a}_\mathbf{k} ^{\dagger} \hat{a}_\mathbf{k},\\
 \hat{H}_\text{I}&= \sum_{j,\mathbf{k}}  g_\mathbf{k} ( e^{{\rm i} \mathbf{k} \cdot\mathbf{r}_j}\hat{O}_{j}^{+} \hat{a}_\mathbf{k} +\text{H.c.} ),
\end{split}\end{align}
where $\hat{O}_{j}$ and $\omega_0$ are the annihilation operators and frequency of the $j$th quantum system located at $\mathbf{r}_j$, $\hat{a}^{\dagger}_\mathbf{k}$ and $\hat{a}_\mathbf{k}$ are the creation and annihilation operators of the environmental $\mathbf{k}$th mode with frequency $\omega_k$, and $g_\mathbf{k}$ is the coupling strength between the systems and the environment. Our system can be two TLSs when $\hat{O}=\hat{\sigma}^-$ \cite{Gonzalez-Tudela2011} or two harmonic oscillators when $\hat{O}=\hat{b}$ \cite{PhysRevA.76.042127, Zell2009}. Here the rotating-wave approximation is used in $\hat{H}_\text{I}$, which is valid in the weak-coupling limit. Under this approximation, the total excitation number $\hat{\mathcal{N}}=\sum_{j,\mathbf{k}} (\hat{O}_{j}^{\dag}\hat{O}_j+\hat{a}^{\dagger}_\mathbf{k}\hat{a}_\mathbf{k})$ of the system is conserved since $[\hat{\mathcal{N}}, \hat{H}]=0$. The Hilbert space of the whole system is thus divided into independent subspaces with definite excitation number $\mathcal{N}$.

Previously, it was found that when the two systems are placed in the same position, there is the DFS \cite{Lidar1998} $|\Psi_\text{DFS}\rangle =\frac{1}{\sqrt{2}}(\hat{O}_1^\dag-\hat{O}_2^\dag)|0,0\rangle$ with $|0\rangle$ being the ground state of the systems due to $\hat{H}_\text{I} |\Psi_\text{DFS}\rangle=0$. This DFS physically originates from the permutation symmetry of the quantum systems \cite{Lidar2003}. We here are interested in exploring whether the DFS still exists when the two systems are placed in different positions such that the permutation symmetry is broken.

\section{Decoherence free state}\label{DFSBMNBM}
\subsection{DFS under BMA}
To describe the DFS of two spatially separated quantum systems influenced by the common zero-temperature environment, we consider first its decoherence dynamics under the BMA. The master equation reads \cite{Scully1997}
\begin{eqnarray}\label{MasterEquation}
& &\dot{\rho}(t) =-{\rm i}[\sum_{i} (\omega_{0}+\Omega_{ii}) \hat{O}^{\dag}_{i} \hat{O}_{i} + (\Omega_{12}\hat{O}^{\dag}_{1} \hat{O} _{2}+\text{H.c.}), \rho(t)] \nonumber \\
& &~~~~+\sum_{i,j} \frac{\gamma_{ij}}{2}\Big[2\hat{O}_{j} \rho(t) \hat{O}^{\dag}_{i} -\{\hat{O}^{\dag}_{i} \hat{O}_{j}, \rho(t)\}\Big]\equiv \check{\mathcal{L}}\rho(t),
\end{eqnarray}
where $\rho(t)$ is the reduced density matrix of the systems, \[\Omega_{ij}=\mathcal{P} \sum_\mathbf{k} \frac{g^{2}_\mathbf{k} e^{{\rm i} \mathbf{k} \cdot(\mathbf{r}_i-\mathbf{r}_j)} }{\omega_k-\omega_0}\]
 with $\mathcal{P}$ the Cauchy principal value, $i=j$ denoting the frequency shift, and $i\neq j$ denoting the dipole-dipole interaction strength induced by the environment, the decay rate reads
\begin{eqnarray}
\gamma_{ij}= 2 \pi \sum_\mathbf{k} g^{2}_\mathbf{k} e^{{\rm i} \mathbf{k} \cdot(\mathbf{r}_{i}-\mathbf{r}_{j})} \delta(\omega_0-\omega_k).\label{DecayRate}
\end{eqnarray}
Generally, $\Omega_{12}$ is real due to the parity symmetry $\Omega_{12}=\Omega_{21}=\Omega_{12}^* $. It can be seen from Eq. (\ref{MasterEquation}) that the environment can not only induce individual spontaneous emission $\gamma_{jj}$ and frequency shift $\Omega_{jj}$ to each system, but also induce the correlated spontaneous emission $\gamma_{12}=\gamma_{21}$ and coherent dipole-dipole interaction $\Omega_{12}$ between the two quantum systems by the exchange of virtual photons.

When the decay rates $\gamma_{ij}$ satisfy $\gamma_{12} = \gamma_{21}=\pm \gamma_{11}=\pm\gamma_{22}$, there is a DFS
\begin{equation}\label{DFSMatrixM}
\rho_\text{DFS}=|\Psi_\text{DFS}\rangle_\mp \langle \Psi_\text{DFS}|
\end{equation}
with $|\Psi_\text{DFS}\rangle_{\mp}=\frac{1} {\sqrt{2}}(\hat{O}^{\dag}_{1} \mp \hat{O}^{\dag}_{2}) |0, 0\rangle$ due to $\check{\mathcal{L}}\rho_\text{DFS}=0$. Combined with Eq. (\ref{DecayRate}), the explicit criterion for the presence of the DFS (\ref{DFSMatrixM}) is
\begin{equation}\label{Criteria}
\mathbf{k}(\omega_0)\cdot\mathbf{R}=l \pi, ~l \in Z,
\end{equation}with $\mathbf{R}=\mathbf{r}_1-\mathbf{r}_2$ being the relative coordinate of the quantum systems.
The sign in Eq. (\ref{DFSMatrixM}) is ``$-$'' (``$+$'') when $l$ is even (odd). Equation (\ref{Criteria}) illustrates that, given the direction of $\mathbf{R}$, the existence of the DFS requires that all the degenerate wave vectors $\mathbf{k}$ with the same frequency $\omega_0$ must satisfy Eq. (\ref{Criteria}) simultaneously. It strongly limits the existence of the DFS in a multidegenerate environment as in two- and three-dimensional cases, where the degeneracy of $\mathbf{k}$ is generally infinite. This condition is possible only for the one-dimensional-environment case. For example, when the one-dimensional environment is formed by the electromagnetic field, the wave vectors for $\omega_0$ can only take $\pm k$. If $k$ satisfies Eq. (\ref{Criteria}), then $-k$ satisfies it naturally. This explains well why all of the works \cite{Gonzalez-Tudela2011, Wolf2011, Fogarty2013, Taketani2014, Fang2015} on the DFS are in one-dimensional environments.

\subsection{Exact DFS beyond BMA }
The above Markovian theory reveals that the DFS for distant quantum systems only exists in the one-dimensional environment case. A natural question is whether it is still valid in the non-Markovian dynamics. The exploration to this issue is meaningful because the non-Markovian effect is non-negligible in a one-dimensional environment, especially for composite quantum systems. Besides the weak system-environment coupling, the validity of the BMA also requires that the environmental correlation timescale is much shorter than the characteristic time of the system. For the composite quantum system as considered in Eq. (\ref{Hamiltonian}), a new timescale characterizing the communication between the subsystems via the common environment is involved. When this timescale is comparable with the
environmental correlation time, the non-Markovian effect would dominate the dynamics even in the weak-coupling limit. This non-Markovian effect is especially important in a one-dimensional environment \cite{Zheng2013, Addis2013,Gonzalez-Ballestero2013}.

Based on the observation that the DFS must be a system-reduced state of the whole-system eigenstate, only under which it is unchanged by the action of $\hat{H}$, we here calculate the exact eigenstate of Eq. (\ref{Hamiltonian}). The DFS derived in this way can efficiently avoid the BMA used in the preceding section. The eigenstate in the single-excitation subspace can be expanded as
$|\psi \rangle =[\sum_{i=1}^2 c_i\hat{O}_i^\dag+\sum_\mathbf{k} d_\mathbf{k}\hat{a}^\dag_\mathbf{k}]|0,0,\{0_\mathbf{k}\} \rangle$, where $|\{0_\mathbf{k}\}\rangle$ denotes the environmental vacuum state. From the Schr\"{o}dinger equation, we can obtain ($i\neq j$)
\begin{eqnarray}\label{EigenEquation}
 c_i\Big(E-\omega_0-\sum_\mathbf{k} \frac{g_\mathbf{k}^2}{E-\omega_{k}}\Big) = c_j\sum_\mathbf{k} \frac{g_\mathbf{k}^{2}e^{{\rm i} \mathbf{k} \cdot(\mathbf{r}_i-\mathbf{r}_j)}}{E-\omega_{k}} ,
\end{eqnarray}
and \[d_\mathbf{k}=g_\mathbf{k}\sum_{j=1,2}{c_j e^{-{\rm i}\mathbf{k}\cdot\mathbf{r}_j}\over E-\omega_k}\] where $E$ is the eigenenergy \cite{Tong2010, Chen2015}. After eliminating $c_1$ and $c_2$, we have $E$ satisfying
\begin{equation}\label{EigenValue0}
  E=\omega_0+\sum_\mathbf{k} \frac{g_\mathbf{k}^2[1\pm \cos (\mathbf{k}\cdot \mathbf{R}) ]}{E-\omega_{k}},
\end{equation} where we used the environmental spatial reflection symmetry, i.e., the modes $\pm \mathbf{k}$ are degenerate in their $\omega_k$ and $g_\mathbf{k}$, has been used. Substituting Eq. (\ref{EigenValue0}) into Eq. (\ref{EigenEquation}) and using again the spatial reflection symmetry, we obtain $c_1=\pm c_2\equiv C/\sqrt{2}$. Absent in the single-quantum-system case \cite{Tong2010, Miyamoto2005}, the cosine term in Eq. (\ref{EigenValue0}) manifests the interference of the two interaction channels of the quantum systems with the common environment. As seen in the following, it is just this interference term which produces the BIC in our bipartite quantum systems.

In the infinite limit of the environmental modes, there is an integration identity for any function $f(\mathbf{k})$:
\begin{equation}\label{CCPV}
\sum_\mathbf{k} \frac{ f(\mathbf{k})}{E-\omega_{k}}=\mathcal{P} \sum_\mathbf{k} \frac{ f(\mathbf{k}) }{E-\omega_{k}}-{\rm i} \pi \sum_\mathbf{k} f(\mathbf{k}) \delta (E-\omega_k).
\end{equation}
The imaginary part in Eq. (\ref{CCPV}) entering into the eigenenergy $E$ contributes to the dynamics a damping rate. Using this identity in Eq. (\ref{EigenValue0}), we can conclude that, to ensure the existence of the DFS, this imaginary part must vanish for one eigenenergy $E_0$ of Eq. (\ref{EigenValue0}), i.e.,
\begin{equation} \label{CriteriaNonMarkovian}
1\pm\cos[\mathbf{k}(E_0)\cdot\mathbf{R}]=0  \Rightarrow\mathbf{k}(E_0)\cdot\mathbf{R}=l\pi,~l\in Z.
\end{equation}
This criterion is almost the same as Eq. (\ref{Criteria}) under the BMA except that the argument $E_0$ differs from $\omega_0$. The eigenstate with the real $E_0$ under Eq. (\ref{CriteriaNonMarkovian}) is an isolated bound state, while other ones with Eq. (\ref{CriteriaNonMarkovian}) unsatisfied are called resonant states playing a significant role in Fano effect \cite{Fano1961, Miroshnichenko2010}. The eigenenergy of the bound state falls in the environmental continuous energyband \cite{PhysRevA.11.446}, it thus is a BIC. Equation (\ref{CriteriaNonMarkovian}) describes the destructive interference of the two interaction channels of the quantum systems with the environment. The BIC here has a close analogy with that predicted in molecule systems \cite{PhysRevA.32.3231}.  After tracing over the environmental degrees of freedom from the BIC, we obtain the exact DFS as
\begin{equation} \label{DFSMatrixNonMarkovian}
\rho_\text{DFS}=|C|^{2} |\Psi_\text{DFS}\rangle_{\pm} \langle \Psi_\text{DFS} | +(1-|C|^{2}) |0,0\rangle \langle 0,0|,
\end{equation}
where $\pm$ depend on the parity of $l$ in Eq. (\ref{CriteriaNonMarkovian}).
It can be seen that, different from the result (\ref{DFSMatrixM}) under the BMA, the exact DFS is a classical mixture of $|\Psi_\text{DFS}\rangle _{\pm}$ and $|0,0\rangle$.

The above analysis reveal that, to make the DFS (\ref{DFSMatrixNonMarkovian}) exist, Eq. (\ref{CriteriaNonMarkovian}) must be satisfied simultaneously by all the degenerate modes $\mathbf{k}$ having the common eigenenergy $E_0$. This again is possible for the one-dimensional-environment case when $\mathbf{k}$ is either parallel or antiparallel to $\mathbf{R}$. So we have $|d_{k}|^{2}=g^{2}_{k}[1 \pm \cos(k R)]|C|^2/(E_0-\omega_{k})^2$. With the help of the normalization condition $|C|^2+\sum_k|d_k|^2=1$, we obtain
\begin{equation}
|C|^2=\Big[1+ \int d\omega {J(\omega)[1\pm \cos (k(\omega)R)]\over (E_0-\omega)^2}\Big]^{-1},\label{absc}
\end{equation}where the summation over $k$ has been replaced by the integration over $\omega$ in the infinite limit of the environmental modes and $J(\omega)=\sum_kg_{k}^2\delta(\omega-\omega_k)$ is the environmental spectral density. Acting as the weight of $|\Psi_\text{DFS}\rangle_\pm$ in $\rho_\text{DFS}$ [Eq. (\ref{DFSMatrixNonMarkovian})], $|C|^2$ determines the entanglement available in the DFS. We notice that the main contribution to the integration in Eq. \eqref{absc} comes from the modes with $\omega \approx E_0$. Making a Taylor expansion near $E_0$, it can be recast into
\begin{eqnarray}
|C|^2&\approx&\Big[1+ J(E_0)\int d\delta {[1-\cos (\alpha R\delta)]\over \delta^2}\Big]^{-1}\nonumber \\
&=&[1+J(E_0)\pi|\alpha| R]^{-1}\approx [J(E_0) \pi|\alpha| R]^{-1},
\end{eqnarray}where $\delta=\omega-E_0$, $\alpha={\partial_\omega k(\omega)|_{\omega=E_0}}$, and Eq. (\ref{CriteriaNonMarkovian}) have been used.
It demonstrates that the weight of $|\Psi_\text{DFS}\rangle_\pm$ in the exact DFS scales as $1/R$ with increasing distance between the two quantum systems. In practice, one generally is interested in realizing distant entanglement distribution by using the DFS \cite{Gonzalez-Tudela2011, Wolf2011, Fogarty2013, Taketani2014}. Our exact result implies that the available entanglement decays in power law as $1/R$ with the increase of the distance between quantum systems. This sets a practical bound on the performance of the scheme.

\section{Illustrational examples}\label{illsmodel}
\begin{figure}[tbp]
  \includegraphics[width=1.\columnwidth]{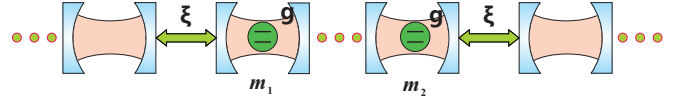}
  \caption{Schematic diagram of a one-dimension nearest-neighbor coupled cavity array with two TLSs embedded in cavities $m_1$th and $m_2$th, respectively. The distance between two nearest-neighbor cavities is $x_0$. }\label{Scheme}
\end{figure}

We first consider two TLSs embedded in a one-dimensional environment formed by a nearest-neighbor coupled-cavity array (see Fig. \ref{Scheme}) with the Hamiltonian
\begin{eqnarray} \label{HamiltonianC}
 \hat{H}_\text{E}= \sum_{j=1}^{N} [ \omega_c \hat{a}_{j} ^{\dagger} \hat{a}_j+\xi (\hat{a}_{j+1} ^{\dagger} \hat{a}_j +\text{H.c.})],
\end{eqnarray}
where $\hat{a}_{j}$ and $\hat{a}^\dag_{j}$ are the annihilation and creation operators of the $j$th cavity with frequency $\omega_{c}$, and $\xi$ is the coupling strength between the nearest-neighbor cavities separated in distance $x_0$. By a Fourier transformation $\hat{a}_{j}=\sum_k\hat{a}_{k} e^{{\rm i} k j x_0}/\sqrt{N}$, Eq. (\ref{HamiltonianC}) is recast into $\hat{H}_\text{E}= \sum_{k} \omega_{k} \hat{a}^{\dagger}_{k} \hat{a}_{k}$ with dispersion relation $\omega_k=\omega_c+2\xi \cos(k x_0)$. Thus the coupled cavity array defines an environment with finite bandwidth $4\xi$ centered at its eigenmode $\omega_c$. The two TLSs are embedded in cavities $m_1$th and $m_2$th, respectively. The interactions are
\begin{equation}
\hat{H}_\text{I}=g\sum_{j=1,2}(\hat{\sigma}_j^+\hat{a}_{m_j}+\text{H.c.}),
\end{equation}which, in the Fourier space, takes the form $\hat{H}_\text{I}=(g/\sqrt{N})\sum_{j=1,2}\sum_k[e^{{\rm i}km_jx_0}\hat{a}_k\hat{\sigma}^+_j+\text{H.c.}]$.
It was previously found that if the TLS frequency falls in the band-gap regime of the environmental spectrum, i.e., $\omega_0<\omega_c-2\xi$ or $\omega_0>\omega_c+2\xi$, a bound state in the band-gap forms \cite{Miyamoto2005, PhysRevA.87.052139}, which has been found to play a constructive role in decoherence suppression \cite{Jorgensen2011, Leistikow2011}, entanglement trapping \cite{Tong2010}, and entanglement generation \cite{Shahmoon2013}. Here we exclude this situation from consideration and concentrate on the potentially formed BIC. Such a bound state has been found in quantum Hall insulators \cite{Yang2013} and optical waveguide array structure \cite{PhysRevLett.100.183902, PhysRevLett.107.183901}.

We see from the dispersion relation that there is a two-fold degeneracy with $\pm k$ for one explicit $\omega_k$. It can be calculated exactly from Eq. (\ref{EigenValue0}) that the eigenenergy of the DFS (\ref{DFSMatrixNonMarkovian}) is $E_0=\omega_0$ (see Appendix \ref{appeign}), which in turn reduces criterion (\ref{CriteriaNonMarkovian}) for forming the DFS  (\ref{DFSMatrixNonMarkovian}) to
\begin{equation} \label{CriteriaCC}
\Delta m\arccos(\frac{\omega_{0}-\omega_{c}}{2\xi})=l \pi,
\end{equation} and the weight (\ref{absc}) to
\begin{eqnarray}\label{EVRFCC}
|C|^2&=&\big[1 +\frac{g^{2}\Delta m}{4 \xi ^2-(\omega_c-\omega_{0})^2} \big]^{-1},
\end{eqnarray}with $\Delta m=m_1-m_2$. The obtained result $E_0=\omega_0$ confirms that the DFS is a BIC because $\omega_0$ is within the environmental energyband. Equation (\ref{EVRFCC}) reveals that $|C|^2$ scales as $1/\Delta m$ with the increase of the TLS distance, which is consistent with the conclusion in last section.

To check whether the BIC is free of decoherence, we resort to the dynamics under the initial condition $|\Phi(0)\rangle=|1,0,\{0_{k}\}\rangle$. Its evolved state takes the form
\begin{equation}\label{tmevo}
|\Phi(t)\rangle=[\sum_{i=1,2}\alpha_i(t)\hat{\sigma}^+_i+\sum_k\beta_k(t)\hat{a}^\dag_\mathbf{k} ]|0,0,\{0_{k}\} \rangle ,
\end{equation}where $\alpha_i(t)$ are governed by
\begin{equation}\label{SchrodingerEquation}
\dot{\alpha}_{i}(t)+{\rm i} \omega_0 \alpha_{i}(t)+\sum_{j=1,2}\int^{t}_{0} d\tau f_{ij}(t-\tau)\alpha_{j}(\tau)=0.
\end{equation}
with $f_{ij}=(g^{2}/N)\sum_{k} e^{-{\rm i} \omega_{k} (t-\tau)+{\rm i} k (m_{i}-m_{j})x_0 }$. The convolution in Eq. (\ref{SchrodingerEquation}) keeps all the non-Markovian
effect induced by the backactions of the memory environment. By the Laplace transform $\tilde {F}(s)=\int_0^\infty e^{-st}F(t)dt$, it is straightforward to show that
\begin{eqnarray}
\tilde{\alpha}_1(s)&=&\sum_{j=0,1}\frac{1/2}{s+\mathrm{i}\omega _{0}+\frac{g^{2}}{N}\sum_{k}\frac{%
1+(-1)^j\cos (k\Delta mx_{0})}{s+\mathrm{i}\omega _{k}}},\label{a1s}\\
\tilde{\alpha}_2(s)&=&-\frac{(g^{2}/N)\sum_{k}e^{\mathrm{i}k\Delta mx_{0}}(s+\mathrm{i}\omega _{k})^{-1}}{s+%
\mathrm{i}\omega _{0}+\frac{g^{2}}{N}\sum_{k}(s+\mathrm{i}\omega _{k})^{-1}}\tilde{\alpha}_{1}(s),\label{a2s}
\end{eqnarray}
Setting $s=-{\rm i}E$, one can find that the pole of Eq. (\ref{a1s}) satisfies Eq. (\ref{EigenValue0}) and thus corresponds exactly to the eigenenergy of the BIC. There is no further pole in Eq. (\ref{a2s}). Using the residue theorem, we readily have
\begin{eqnarray}
\alpha_j(t)&=&Z_{j}e^{-{\rm i}\omega _{0}t}+\int_{{\rm i}\epsilon -\infty}^{ \prime{\rm i}\epsilon +\infty} \frac{dE}{2\pi }\tilde{\alpha}_{j}(-{\rm i}E)e^{-{\rm i}Et},\label{astt}
\end{eqnarray}where the first term with residue $Z_j$ is contributed from the BIC with eigenenergy $\omega_0$, the second term contains all the contributions from the continuous energyband, and the prime in the integration represents the integration region excluding the eigenenergy of the BIC. It is interesting to find that the two residues take exactly as
\begin{equation}
Z_1=\pm Z_2=|C^2|/2.
\end{equation}Oscillating with time in continuously changing frequencies, the second term in Eq. (\ref{astt}) behaves as a decay and approaches zero in the long-time limit due to the out-of-phase interference. Therefore, the steady state of our system after tracing over the environmental degrees of freedom from $|\Phi(\infty)\rangle$ is
\begin{equation}\label{rhoinfty}
\rho(\infty)={|C|^2\over 2}\rho_\text{DFS}+(1-{|C|^2\over 2})|0, 0\rangle\langle0, 0|,
\end{equation}which demonstrates the dominate role of the formed $\rho_\text{DFS}$ in the long-time steady state.
On the contrary, if no BIC and DFS formed, then $\rho(\infty)=|0, 0\rangle\langle 0, 0|$. The results verify analytically from the point of view of the dynamics the validity of our expectation that the reduced state of the formed BIC is a DFS of our decoherent system.

\begin{figure}[tbp]
  \includegraphics[width=1.\columnwidth]{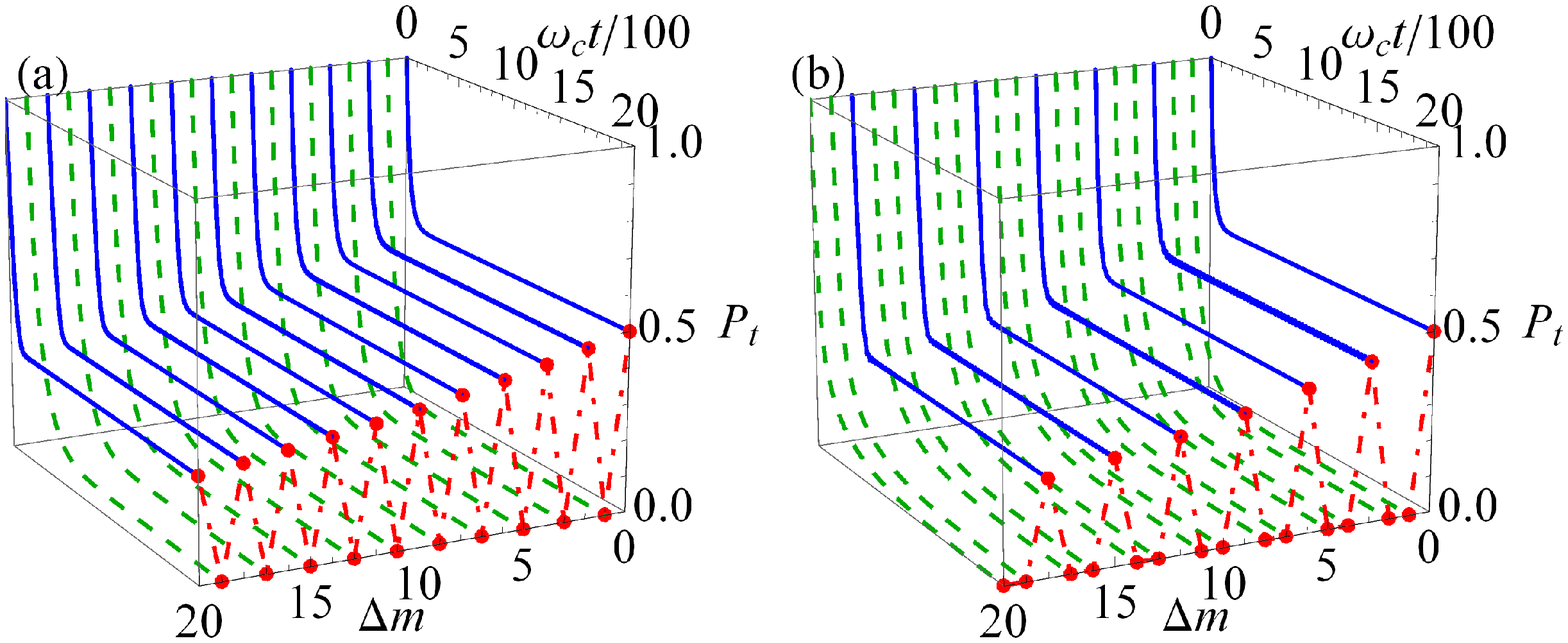}\\
  \vspace{0.5cm}
  \includegraphics[width=1.\columnwidth]{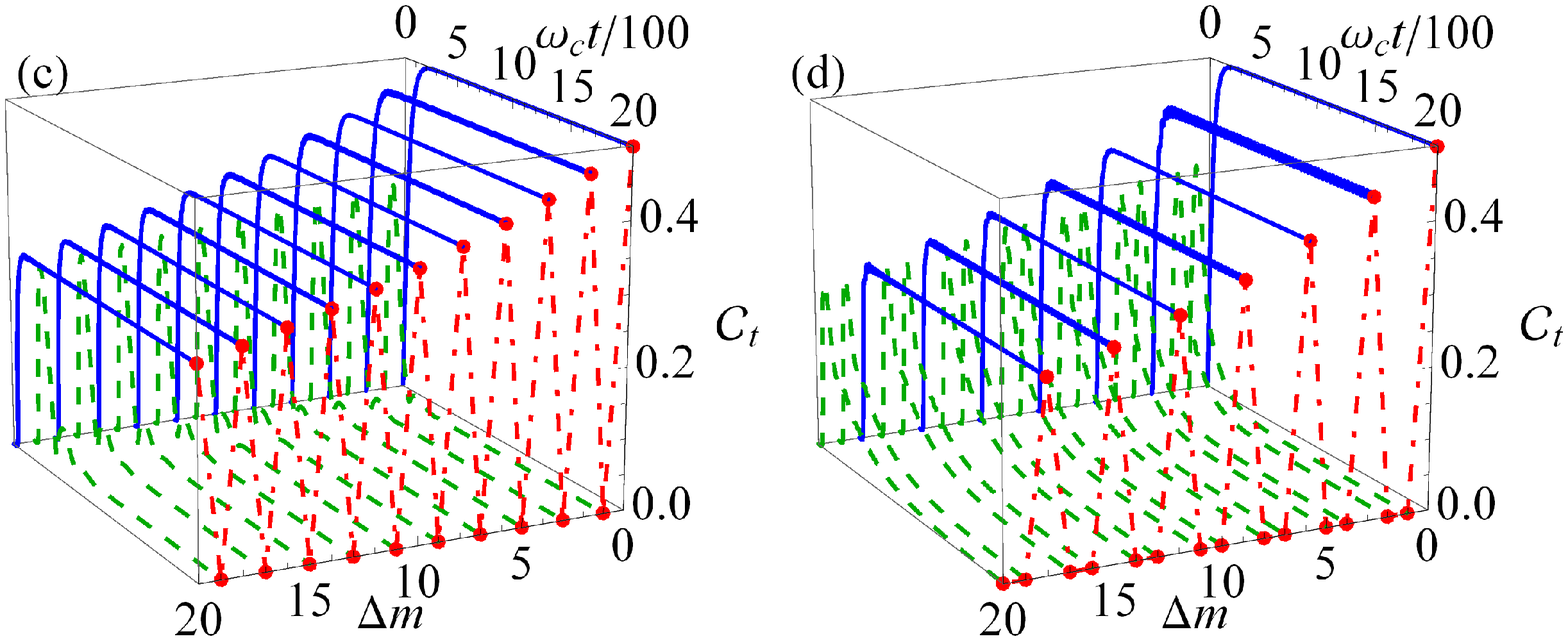}
  \caption{Time evolution of the TLS excited-state population $P_{t}$ in panels (a) and (b) and concurrence $\mathcal{C}_t$ in panels (c) and (d) as the change of the TLS separation $\Delta m$ when $\omega_{0}=1.0\omega_{c}$ in panels (a) and (c) and $1.2\omega_{c}$ in panels (b) and (d) obtained by numerically solving Eqs. (\ref{SchrodingerEquation}). The blue solid and green dashed lines denote the cases with and without the DFS formed, respectively. The dots connected by the red dotdashed lines plot the results analytically evaluated from Eq. (\ref{EVRFCC}). Other parameters are $\xi=0.2\omega_c$, $g=0.05\omega_c$, and $N=1201$.} \label{dec-free}
\end{figure}

We plot the time evolution of the total excited-state population $P_t=\sum_{i=1,2}|\alpha_i(t)|^2$ in different TLS separation $\Delta m$ when $\omega_0=\omega_c$ in Fig. \ref{dec-free}(a) and $1.2\omega_c$ in Fig. \ref{dec-free}(b) calculated by numerically solving Eq. (\ref{SchrodingerEquation}). Equation (\ref{CriteriaCC}) indicates that the DFS is formed when $\Delta m$ is an even number for $\omega_0=\omega_c$. Figure \ref{dec-free}(a) indeed shows that $P_t$ tends to a finite value when $\Delta m$ is an even number, and decays to zero whenever $\Delta m$ is an odd number. The preserved steady-state population matches well with $\text{Tr}[\rho(\infty)\sum_{j=1,2}\hat{\sigma}_j^+\hat{\sigma}_j^-]=|C|^4/2$ calculated analytically from Eq. (\ref{EVRFCC}), which verifies unambiguously that the initial state evolves exclusively to the $\rho(\infty)$ obatined in  Eq. (\ref{rhoinfty}). When $\omega_{0}=1.2\omega_{c}$, the DFS is formed when $\Delta m=3n$ with $n$ an integer. As confirmed by Fig. \ref{dec-free}(b), $P_t$ approaches $|C|^4/2$ when $\Delta m=3n$ and decays to zero in other cases  without any exception. The exact correspondence between the analytical results $|C|^4/2$ and the numerical dynamics testifies the distinguished role played by the formed DFS in the steady-state behavior.

To further verify the validity of Eq. (\ref{rhoinfty}), we plot the evolution of the entanglement between the TLSs in Figs. \ref{dec-free}(c) and \ref{dec-free}(d). The entanglement is quantified by concurrence \cite{PhysRevLett.80.2245}, which for the state \eqref{tmevo} is $\mathcal{C}_t=2|\alpha_1 (t) \alpha_2 (t)|$. We can find that the parameter regimes in Figs. \ref{dec-free}(a) and \ref{dec-free}(b) where a nonzero $P_t$ is achieved match exactly well with the regimes in Figs. \ref{dec-free}(c) and \ref{dec-free}(d) where a finite concurrence is obtained. The concurrence approaches the analytical value $|C|^4/2$, which is just the concurrence calculated from the steady state (\ref{rhoinfty}). It demonstrates well the distinguished role played by the formed DFS in the dynamics and steady-state behavior.

 \begin{figure}[tbp]
  \includegraphics[width=1.\columnwidth]{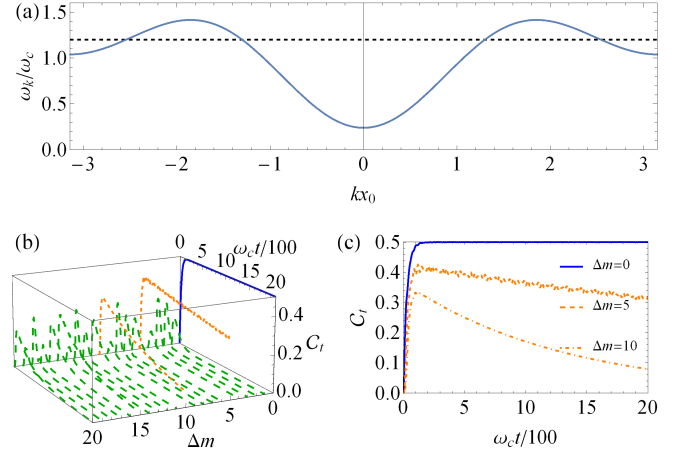}
  \caption{(a): Environmental dispersion relation reveals that a four-fold degeneracy exists when $\omega_{k}>1.04\omega_c$. (b) and (c): Time evolution of $\mathcal{C}_{t}$ in different TLS separation $\Delta m$ when the next-nearest-neighbor hopping of the cavity array is considered. The blue solid and green dashed lines denote the cases with and without the DFS formed, respectively. The orange dotted lines show $\mathcal{C}_t$, although dramatically is slowed down, finally decays to zero. The parameter $\xi'=0.9\xi$ and the others are the same as in Fig. \ref{dec-free}(b). }\label{nextnearest}
\end{figure}

On the other hand, if the environment is not two-fold degeneracy, then the criterion (\ref{CriteriaNonMarkovian}) for forming the DFS is hard to be satisfied even for the one-dimensional environment case. To verify this, we next consider another situation where the next-nearest-neighbour coupling of the cavity array is involved. Then Eq. (\ref{HamiltonianC}) is recast into
\begin{equation}\label{nextnearestH}
 \hat{H}'_\text{E}= \sum_{j}[ \omega_c \hat{a}_{j} ^{\dagger} \hat{a}_j+(\xi \hat{a}_{j+1} ^{\dagger} \hat{a}_j +\xi' \hat{a}_{j+2} ^{\dagger} \hat{a}_{j} + \text{H.c.})],
\end{equation}where $\xi'$ is the next-nearest-neighbor hopping rate. The dispersion relation is derived to be
\begin{equation}\label{DispersionRelationNN}
\omega_{k}=\omega_{c}+2\xi \cos(k x_0)+2\xi' \cos(2 k x_0).
\end{equation}
Compared with the two-fold degeneracy in the nearest-neighbor hopping case, the modes here can take four-fold degeneracy $\pm k_1$ and $\pm k_2$ [see Fig. \ref{nextnearest}(a)]. The formation of the DFS require that the criterion (\ref{CriteriaNonMarkovian}) should be satisfied for $\pm k_{1}$ and $\pm k_{2}$ simultaneously. We numerically calculate the dynamics and plot the $P_t$ obtained in Fig. \ref{nextnearest}(b). It shows that the DFS previously formed in Fig. \ref{dec-free}(b) disappears with the next-nearest-neighbor hopping considered. Thus no stable concurrence can be established in the long-time limit except for the trivial case $\Delta m=0$. Although in some cases the decay of $\mathcal{C}_t$ transiently formed is dramatically slowed down, it decays to zero asymptotically [see Fig. \ref{nextnearest}(c)]. This gives a counterexample to illustrate the validity of criterion (\ref{CriteriaNonMarkovian}).

\section{Conclusions}\label{con}
In summary, we investigated the DFS of two distant quantum systems embedded in a common environment. Going from a general model of dissipative systems, we derived the criterion for forming the DFS for both of the Markovian and non-Markovian decoherence dynamics. It is interesting to find that the DFS may be formed only in one-dimensional environment case. We have also revealed that the exact DFS for the non-Markovian dynamics is a reduced density matrix of the so-called BIC of the total system, which consists of a classical mixture of the maximally entangled state $|\Psi\rangle_\pm$ and $|0,0\rangle$. The weight of the former scales as $1/R$ with the increase of the system distance, which sets a bound on distributing entanglement over distant quantum systems via the common environment. The exact dynamics of two TLSs embedded in two types of coupled cavity array as the common environment are studied explicitly, which verifies our prediction on the DFS. Our scheme supplies an implementation of Friedrich and Wintgen's idea on the realization of the  mathematically curious BIC in explicit quantum optical system \cite{PhysRevA.11.446}. By giving insight into the physical nature of the DFS, our exact DFS result is expected to be helpful to interpret the environment induced entanglement between two quantum systems.

\section*{Acknowledgments}
This work was supported by the Specialized Research Fund for the Doctoral Program of Higher Education, by the Program for New Century Excellent Talents in University, and by the National Natural Science Foundation of China (Grant Nos. 11175072 and 11474139).

\appendix
\section{Derivation of the eigenvalue of the bound state in the continuum}\label{appeign}
The eigenvalue of the BIC for the coupled cavity array environment satisfies
\begin{equation}\label{egq}
E_0=\omega_{0}+{g^{2}\over N}\sum_{k}\frac{1\pm\cos(kx_0\Delta m)}{E_0-\omega_{k}},
\end{equation}where $\pm$ are determined by the separation $\Delta m$ according to Eq. (\ref{CriteriaCC}). With the dispersion relation $\omega_{k}=\omega_{c}+2\xi\cos (kx_{0})$, Eq. (\ref{egq}) in the continuous limit of the environmental modes is recast into
\begin{equation}
E_0=\omega_{0}+\frac{g^2x_0}{2\pi}\int_0^{2\pi\over x_0} dk\frac{1\pm\cos(kx_0\Delta m)}{E_0-\omega_{c}-2\xi\cos(kx_{0})}.
\end{equation}
Setting $z=e^{{\rm i}kx_{0}}$, we have
\begin{widetext}
\begin{eqnarray}
E_0 & = & \omega_{0}+\frac{ g^{2}}{2\pi {\rm i}}\ointctrclockwise _{z=1}dz\frac{1\pm\frac{1}{2}(z^{\Delta m}+z^{-\Delta m})}{z(E_0-\omega_{c})-\xi(z^2+1)}\nonumber \\
 & = & \omega_{0}-\frac{g^{2}}{2\xi}\ointctrclockwise _{z=1}\frac{dz}{2\pi{\rm i}}\frac{1\pm z^{\Delta m}}{(z-a-{\rm i}\epsilon)(z-a^{\ast}-{\rm i}\epsilon)}-\frac{g^{2}}{2\xi}\ointctrclockwise _{z=1}(-z^{2})\frac{dz^{-1}}{2\pi {\rm i}}\frac{1\pm z^{-\Delta m}}{[z-a-{\rm i}\epsilon][z-a^{\ast}-{\rm i}\epsilon]}\nonumber\\
 & = & \omega_{0}-\frac{g^{2}}{2\xi}\ointctrclockwise _{z=1}\frac{dz}{2\pi{\rm i}}\frac{1\pm z^{\Delta m}}{(z-a-{\rm i}\epsilon)(z-a^{\ast}-{\rm i}\epsilon)}+\frac{g^{2}}{2\xi}\varointclockwise_{z'=1}\frac{dz^{'}}{2\pi {\rm i}}\frac{1\pm z'^{\Delta m}}{[z'-(a+{\rm i}\epsilon)^{-1}][z'-(a^{\ast}+{\rm i}\epsilon)^{-1}]},\label{integ}
\end{eqnarray}\end{widetext}
where $a$ and $a^{\ast}$ are the solutions of the equation $z^{2}-z(E_0-\omega_{c})/\xi+1=0$ and satisfy $a^{\ast}a=1$ and $\epsilon$ is an infinitesimal positive value. Here an integration relation $\ointctrclockwise_{z=1}dz=\varointclockwise_{z^{-1}=1}dz^{-1}$ has been used. There are two singularities for each of the integrations in Eq. (\ref{integ}). However, only one of them falls within the circle $z=1$ and $z'=1$, respectively. According to the residue theorem, we can evaluate the integrations as
\begin{eqnarray}
E_0 & = & \omega_{0}-\frac{g^{2}}{2\xi}\Big[\frac{1\pm a^{\Delta m}}{a-a^{\ast}}+\frac{1\pm (a^{\ast})^{-\Delta m}}{(a^{\ast})^{-1}-a^{-1}}\Big]\nonumber\\
 & = & \omega_{0}-\frac{g^{2}(1\pm a^{\Delta m})}{\xi(a-a^{\ast})}.
\end{eqnarray}
Because $E_0$, as the the eigenvalue of the BIC, must be real, we readily have $1\pm a^{\Delta m}=0$ and thus $E_0 = \omega_{0}$. Substituting the form of $a={\omega_0-\omega_c\over 2\xi}-{\rm i}\sqrt{1-({\omega_0-\omega_c\over 2\xi})^2}=\exp[-{\rm i}\arccos({\omega_0-\omega_c\over 2\xi})]$ into $1\pm a^{\Delta m}=0$, we readily have
\begin{eqnarray}
\Delta m\arccos({\omega_0-\omega_c\over 2\xi})=l\pi,
\end{eqnarray}which is just the criterion Eq. (\ref{CriteriaCC}) for forming the BIC.

With the similar procedure, the weight $|C|^2$ can also be evaluated as
\begin{eqnarray}
|C|^2 & = & \Big[1+\frac{g^2 x_0}{2\pi}\int_0^{2\pi\over x_0} dk\frac{1\pm \cos(kx_0\Delta m)}{(E_0-\omega_{c}-2\xi\cos(kx_{0}))^{2}}\Big]^{-1}\nonumber \\
 & = & \Big[1\pm \frac{g^{2}\Delta m a^{\Delta m}}{\xi^{2}(a-a^{\ast})^{2}}\Big]^{-1}\nonumber \\
 &=& \Big[1+\frac{g^{2}\Delta m}{4\xi^{2}-(\omega_{0}-\omega_{c})^{2}}\Big]^{-1},
\end{eqnarray}
where $1\pm a^{\Delta m}=0$ has been used.

\bibliography{common-bath}

\end{document}